\documentclass[aip,jcp,notitlepage,twocolumn,reprint,floatfix]{revtex4-2}

\usepackage{graphicx}
\usepackage{dcolumn}
\usepackage{bm}
\usepackage{hyperref}
\usepackage{amsmath}
\usepackage{mathtools}
\usepackage{booktabs}
\usepackage{multirow}
\usepackage{comment}
\usepackage{natbib}
\usepackage{xcolor}
\let\oldtheequation\theequation
\makeatletter
\def\tagform@#1{\maketag@@@{\ignorespaces#1\unskip\@@italiccorr}}
\renewcommand{\theequation}{(\oldtheequation)}
\makeatother

\newcommand{\appref}[1]{\hyperref[#1]{Appendix~\ref{#1}}}
\newcommand{\lit}[1]{Ref.~\onlinecite{#1}}

\begin{document}

\title{Low communication high performance \emph{ab initio} density matrix renormalization group algorithms}

\author{Huanchen Zhai}
\email{hczhai@caltech.edu}
\affiliation{Division of Chemistry and Chemical Engineering, California Institute of Technology, Pasadena, CA 91125, USA}

\author{Garnet Kin-Lic Chan}
\email{gkc1000@gmail.com}
\affiliation{Division of Chemistry and Chemical Engineering, California Institute of Technology, Pasadena, CA 91125, USA}

\date{\today}

\begin{abstract}
There has been recent interest in the deployment of  \textit{ab initio} density matrix renormalization group computations on high performance computing platforms. Here, we introduce a reformulation of the conventional distributed memory \textit{ab initio} DMRG  algorithm that connects it to the conceptually simpler and advantageous sum of sub-Hamiltonians approach. Starting from this framework, we further explore a hierarchy of parallelism strategies, that includes (i) parallelism over the sum of sub-Hamiltonians, (ii) parallelism over sites, (iii) parallelism over normal and complementary operators, (iv) parallelism over symmetry sectors, and (v) parallelism within dense matrix multiplications. We describe how to reduce processor load imbalance and the communication cost of the algorithm to achieve higher efficiencies. We illustrate the performance of our new open-source implementation on a recent benchmark ground-state calculation of benzene in an orbital space of 108 orbitals and 30 electrons, with a bond dimension of up to 6000, and a model of the FeMo cofactor  with 76 orbitals and 113 electrons. The observed parallel scaling from 448 to 2800 CPU cores is nearly ideal.
\end{abstract}

\maketitle

\section{Introduction}

The Density Matrix Renormalization Group (DMRG) algorithm\cite{white1992density, white1993density} is established as a method to obtain highly accurate low-energy eigenstates of \textit{ab initio} quantum chemistry Hamiltonians\cite{white1999ab,mitrushenkov2001quantum,mitrushenkov2003quantum,chan2002highly,legeza2003qc,chan2004state,moritz2006construction,hachmann2006multireference,kurashige2009high,marti2011new,chan2011density,fertitta2014investigation,olivares2015ab}. While multiple techniques can now solve for low-energy eigenstates to high precision in problems that are formally beyond the reach of full configuration interaction~\cite{holmes2016heat,sharma2017semistochastic,booth2009fermion,blunt2017density,eriksen2020ground}, DMRG provides a unique capability to treat problems with a large number of open shells.\cite{kurashige2011second,guo2018communication} Consequently it is particularly useful in active space problems where a large fraction of the orbitals have open shell character, for example, as found in molecular clusters with multiple open-shell transition metal centers.\cite{marti2008density,kurashige2013entangled,kurashige2014complete,chalupsky2014reactivity,sharma2014low,li2019electronic} In many such problems, teasing out the relevant chemistry requires not only a single ground-state energy calculation, but also the characterization of many competing low-energy states. For such applications, improving the scalabity and efficiency of current DMRG implementations is highly desirable. 

Over the last two decades, many different strategies have been proposed to parallelize the DMRG algorithm in quantum chemistry. These include:

(i) Parallelism within dense matrix multiplications\cite{hager2004parallelization,levy2020distributed}. This is a fine-grained parallelism which is effective when the size of the dense matrices is sufficiently large (namely, when a large Matrix Product State (MPS) bond dimension \( M \) is used). It can be implemented simply by linking the code to a parallelized shared-memory math library.

(ii) Parallelism over symmetry sectors,\cite{kurashige2009high,levy2020distributed} which is available when DMRG is implemented with symmetry restrictions. Typically, particle number, total spin or projected spin, and spatial symmetry are used in \textit{ab initio} DMRG implementations.

(iii) Parallelism over normal and complementary operators\cite{chan2004algorithm,wouters2014density}. This is often considered the largest source of parallelism for typical \textit{ab initio} problems.

(iv) Parallelism over a sum of sub-Hamiltonians\cite{chan2016matrix}. This is a coarse-grained parallelism with very low communication cost, and is easy to express in a Matrix Product Operator (MPO) description of DMRG.

(v) Parallelism over sites\cite{stoudenmire2013real}. For a large number of sites, this is an additional source of coarse grained parallelism.
Such an implementation relies on the transformation of the MPS to a form with multiple canonical centers.

Recently, Brabec et. al. reported a non-spin-adapted massively parallel implementation of DMRG for quantum chemistry using strategies (ii) and (iii).\cite{brabec2020massively} We also note promising recent progress in GPU accelerated parallel DMRG algorithms.\cite{nemes2014density,chen2020improved,li2020numerical} However, to the best our knowledge, there has not been an implementation that utilizes all 5 sources of parallelism in a scalable DMRG code
for \textit{ab initio} problems. This may  be partly ascribed to the fact that strategies (iv) and (v) are most conveniently implemented in a DMRG code \cite{li2017spin,keller2015efficient} that is structured using an MPO/MPS formalism,\cite{chan2016matrix,hubig2017generic} while many other efficient \textit{ab initio} DMRG implementations\cite{sharma2012spin,wouters2014chemps2} using strategies (i), (ii) and (iii) are organized around the construction and transformation of renormalized operators.\cite{chan2002highly}

In this work, we first reformulate strategy (iii) for a distributed memory model using the sum of sub-Hamiltonians language. This demonstrates that a low communication version of strategy (iii) can in fact also be viewed as a variant of strategy (iv). This analysis constitutes \autoref{sec:para-op} and \autoref{sec:para-sub}. In \autoref{sec:para-sites}, we discuss how the load-imbalance that arises in strategy (v) can be alleviated via the dynamical determination of connection sites. In \autoref{sec:para-shared-op} to \autoref{sec:para-dense} we briefly introduce the shared memory parallelism strategies (i), (ii) and (iii) used in this work. In \autoref{sec:impl} we briefly discuss a few lower level implementation details. Next, in \autoref{sec:results}, we illustrate the computational performance of our new implementation of parallel DMRG for a recent ground-state  benzene benchmark\cite{eriksen2020ground} in a polarized valence double zeta basis.\cite{dunning1989gaussian}
Although not a correlated or open-shell system that is particularly suited to DMRG, the size of the calculation  serves to illustrate the scalability of our algorithm. For a correlated electron problem with many open shells that is more suited for DMRG, we also consider a model of the FeMo cofactor system\cite{li2019femoco}, and observe that a similar scaling can be achieved. Finally, the conclusions are given in \autoref{sec:conclude}.

\section{Theory}

Rather than reintroduce the DMRG formalism here, we summarize the background theory and notation for the serial DMRG algorithm\cite{white1992density, white1993density} and the SU(2) (spin-adapted) \textit{ab initio} DMRG algorithm \cite{sharma2012spin,wouters2014density,keller2016spin} in \appref{app:dmrg} and \appref{app:qc-dmrg}, respectively. We encourage readers unfamiliar with the standard DMRG algorithm and terminology to first consult these appendices.

\subsection{Parallelism over renormalized operators} \label{sec:para-op}

In most parallel implementations of \textit{ab initio} DMRG, the most important  source of distributed memory parallelism comes from distributing the left-right renormalized operator decomposition of the Hamiltonian, as discussed in \lit{chan2004algorithm}. In this approach, ``normal" and ``complementary" renormalized operators (see \appref{app:qc-dmrg} for definitions) are assigned to different processors according to their orbital indices.

The leading communication cost per sweep in the approach described in \lit{chan2004algorithm} 
is \( O(16 M^2 K^2 \log P_{\mathrm{hamil}}) \) from the blocking step, where \( M \) is the MPS bond dimension, \( K \) is the number of sites, and \( P_{\mathrm{hamil}} \) is the total number of processors (processor cores) at this parallelism level. The sub-leading term in the communication cost is \( O(M^2 K^2 \log P_{\mathrm{hamil}}) \) from the transformation (rotation) step.

In order to achieve better scalability, it is desirable to reduce the communication cost. For this purpose, we note that the leading and sub-leading terms in the communication cost in the above approach mainly come from the accumulation of the \( R_i^{L/R[\frac{1}{2}]} \) operators (defined in \autoref{eq:op-def}). Therefore, the communication cost can be greatly reduced by never accumulating \( R_i^{L/R[\frac{1}{2}]} \). Namely, we can arrange for each processor to compute and store a partial contribution to \( R_i^{L/R[\frac{1}{2}]} \) for all indices \( i \). Compared to the original scheme, this new scheme only needs to communicate when accumulating the wavefunction, with a communication cost of \( O(16M^2 K \log P_{\mathrm{hamil}}) \) per sweep. However, since all partial components of the \( R_i^{L/R[\frac{1}{2}]} \) operators have to enter into the solving (Davidson) step, the computational cost for the solving step increases from \( O(M^3(K^3 + K^2)/P_{\mathrm{hamil}}) \) to \( O(M^3K^3/P_{\mathrm{hamil}} + M^3K^2) \) per sweep. The total disk storage cost also increases from \( O(M^2(K^3 + K^2)) \) to \( O(M^2K^3 + M^2K^2 P_{\mathrm{hamil}}) \). For the typical case where \( P_{\mathrm{hamil}} \not\gg K \), the increase in the subleading term of the storage is not a large concern.

We note that in this new scheme, the communication of renormalized operators is completely removed. In other words, each processor performs blocking, solving, and transformation steps for a part of the Hamiltonian - i.e. a sub-Hamiltonian - independently, and only wavefunctions from the solving step are communicated. This motivates a more general picture where we can develop low communication algorithms that are formulated in terms of sub-Hamiltonians, rather than the left-right decomposition of the Hamiltonian. 

\subsection{Parallelism over sub-Hamiltonians} \label{sec:para-sub}

For this purpose, we write the \textit{ab initio} Hamiltonian \autoref{eq:qc-hamil} as
\begin{equation}
    \hat{H} = \hat{H}^{(1)} + \hat{H}^{(2)} + \cdots + \hat{H}^{(P_{\mathrm{hamil}})}
    \label{eq:sum-h}
\end{equation}
where \( \hat{H}^{(p)} \) is the sub-Hamiltonian assigned to processor \( p \). To describe this assignment, we can write
\begin{multline}
    \hat{H}^{(p)} = \frac{1}{2} \sum_{ij,\sigma}
        \Big[ \operatorname{proc}(p, i) + \operatorname{proc}(p, j) \Big]
        t_{ij,\sigma} \ a_{i\sigma}^\dagger a_{j\sigma} \\
    + \frac{1}{2} \sum_{ijkl, \sigma\sigma'} \operatorname{proc}(p, i, j, k, l)\ 
        v_{ijkl, \sigma\sigma'}\
        a_{i\sigma}^\dagger a_{k\sigma'}^\dagger a_{l\sigma'}a_{j\sigma}
    \label{eq:sub-h}
\end{multline}
where \( \operatorname{proc}(p, \cdots) \)  defines the mapping from orbital indices to processor rank \( p \) (\( p = 1,2,\cdots, P_{\mathrm{hamil}}\)). There is clearly much freedom in choosing the definition of these mappings.

A possible definition of \( \operatorname{proc}(p, \cdots) \) is
\begin{equation}
\begin{aligned}
    \operatorname{proc}(p, i) =&\ \begin{cases}
        1 & p \equiv i \ (\mathrm{mod}\ P_{\mathrm{hamil}}) \\
        0 & \text{otherwise} \end{cases} \\
    \operatorname{proc}(p, i, j) =&\ \begin{cases}
        1 & p \equiv \frac{(j-1)j}{2} + i \ (\mathrm{mod}\ P_{\mathrm{hamil}})\ \text{and}\ i \le j \\
        1 & p \equiv \frac{(i-1)i}{2} + j \ (\mathrm{mod}\ P_{\mathrm{hamil}})\ \text{and}\ i > j \\
        0 & \text{otherwise} \end{cases}
    \label{eq:proc-def}
\end{aligned}
\end{equation}
and \( \operatorname{proc}(p, i, j, k, l) \) has the same value for any permutation of parameters \( i, j, k, l \), namely
\begin{equation}
    \operatorname{proc}(p, i, j, k, l) =
        \operatorname{proc}(p, \text{sorted:\ }i, j, k, l)
\end{equation}
 As discussed above, we can think of a modified version of the normal-complementary operator parallelism as arising from a particular decomposition into sub-Hamiltonians. In particular, 
 the NC renormalized operator partition (\autoref{eq:h-nc}) corresponds to
\begin{equation}
    \operatorname{proc}(p, \text{sorted:\ }i, j, k, l) = \begin{cases}
        \operatorname{proc}(p, j) & j = k \\
        \operatorname{proc}(p, i, j) & \text{otherwise} \end{cases}
    \label{eq:proc-def-nc}
\end{equation}
while the CN renormalized operator partition (\autoref{eq:h-cn}) corresponds to
\begin{equation}
    \operatorname{proc}(p, \text{sorted:\ }i, j, k, l) = \begin{cases}
        \operatorname{proc}(p, j) & j = k \\
        \operatorname{proc}(p, k, l) & \text{otherwise} \end{cases}
    \label{eq:proc-def-cn}
\end{equation}
Here  the notation \( \operatorname{proc}(p, \text{sorted:\ }i, j, k, l) \) means that \( i \le j \le k \le l \).
Based on the above definitions, the symmetry condition \( \operatorname{proc}(p, i, j) = \operatorname{proc}(p, j, i)\) is satisfied. This is important for efficiency, since the operator symmetry conditions used for efficient DMRG algorithms (see \autoref{eq:index-sym}) can still be used on each processor without any communication.

To see how this assignment of Hamiltonian terms gives the correct scaling for multiple processors, we note that in the NC partition in \autoref{eq:h-nc} the summation over two-index operators is over indices in the left block of sites \( L \), which are the small indices \( i, j \) in the tuple \( i, j, k, l \), and thus in \autoref{eq:proc-def-nc} the indices \( i, j \) are used for the processor assignment. For similar reasons, the large indices \( k, l \) are used in the CN partition case. This ensures that the total number of terms in the left-right decomposition of the effective Hamiltonian on each processor is roughly \( O(K^2/P_{\mathrm{hamil}}) \) (if \( P_{\mathrm{hamil}} \not\gg K \)).

It is worth noting that \autoref{eq:sum-h} is in the same spirit as the sum of MPOs formulation first introduced in \lit{chan2016matrix}. This is often considered  a different strategy from the strategy of parallelism over renormalized operators. The two methods indeed have a very different origin and motivation. However, our new formulation of the low communication version of the parallelism over 
renormalized operators establishes a clear connection between the two methods. In addition, we find that this new formulation inherits the most important advantages from both methods:

(i) Low communication time.\cite{chan2016matrix} Since each sub-Hamiltonian can be manipulated completely independently, only the communication of the (small) wavefunction obtained from the DMRG solving step is required.

(ii) Simple implementation. To parallelize a serial \textit{ab initio} DMRG code, one only needs to start with a distributed integral file, where for each processor some integrals \( t_{ij} \) and \( v_{ijkl} \) are set to zero according to \autoref{eq:sub-h}. A single communication step then needs to be added to accumulate wavefunctions from all processors. No other part of the code needs to be changed significantly.

(iii) Compatibility with both the exact renormalized operator and compressed MPO DMRG formalisms. Because the description of our algorithm does not rely on  specific definitions and choices of normal and complementary operators,\cite{white1999ab} one has great freedom to decompose each sub-Hamiltonian.
For example, we have presented examples that correspond to the conventional exact NC or CN renormalized operator partitions\cite{chan2002highly} and their corresponding MPOs, but other MPO representations of the sub-Hamiltonians, including compressed representations\cite{chan2016matrix} can be used. In this work, we will only use exact MPO representations of the sub-Hamiltonians.



(iv) Compatibility with index-symmetry conditions. We note that the previous description of sub-Hamiltonians in \lit{chan2016matrix} was based on splitting the Hamiltonian based on single site-indices. This has the disadvantage that it becomes difficult to use the two-index symmetry conditions \autoref{eq:index-sym} to reduce the computational cost associated with each sub-Hamiltonian in a distributed setting, because
a single site-index based processor assignment can easily assign index-symmetry related operators to different processors. The current two-index based splitting does not have this problem since \( \hat{O}_{ij} \) and \( \hat{O}_{ji} \) are always assigned to the same processor.

(v) Load balance between processors. The two-index based assignment assigns roughly equal amounts of work to different processors, if \( P_{\mathrm{hamil}} \not\gg K \).

(vi) Compatibility with the sum of sub-Hamiltonians and automated MPO construction. In \lit{chan2016matrix}, it was demonstrated that by expressing the \textit{ab initio} Hamiltonian as a sum of \( K \) sub-Hamiltonians, we can work with \( K \) MPOs each with bond dimension \( O(K) \), instead of one MPO with bond dimension \( O(K^2) \). The  advantage is that this captures the primary sparsity within the single MPO representation of the Hamiltonian. This means that a simple MPO construction of the sub-Hamiltonians, which uses only dense matrices, produces the correct serial computational cost of 
\( O(M^3K^3 + M^2K^4) \), rather than the naive (and incorrect) cost of \( O(M^3K^3 + M^2K^5) \) arising from single dense MPO representation of $H$, making the correct scaling of the \textit{ab initio} implementation  very easy to achieve. In particular, this is attractive when combined with various automated MPO construction approaches, which then do not need to implement sparse tensor algebra.\cite{keller2015efficient,chan2016matrix,hubig2017generic,ren2020general} 
The two-index based sum of sub-Hamiltonians retains this attractive feature, but has the further advantage that the computational prefactor   (e.g. from the sub-MPO bond dimensions) is smaller, when compared with the previous one-index decomposition.

In this work, we combine the low communication scheme based on sub-Hamiltonians with a mixed NC/CN partition\cite{kurashige2009high} to achieve high efficiency. The mixed NC/CN partition introduces additional costs for computation and communication  at the middle site of the sweep. These details are discussed in \autoref{sec:mixed}.

\subsection{Parallelism over sites} \label{sec:para-sites}

A more recent approach to coarse-grained parallelism in DMRG is the ``real space parallel DMRG" approach introduced by Stoudenmire and White,\cite{stoudenmire2013real, secular2020parallel} which has been shown to give near ideal scaling in some calculations with model Hamiltonians and very recently for quantum chemistry Hamiltonian.\cite{chen2021real} An implementation of this approach for (non-spin-adapted) quantum chemistry Hamiltonians can also be found in the \textsc{QCMaquis} code.\cite{keller2015efficient}

The approach relies on a representation of the MPS with multiple canonical centers.\cite{stoudenmire2013real}
Each extra canonical center can be introduced by first performing a SVD on the effective wavefunction (given in \autoref{eq:eff-wfn}) at the original canonical center \( k \)
\begin{equation}
    \mathbf{\Psi}[k]^{\mathrm{eff}} = \mathbf{L}[k]
        \mathbf{S}[k] \mathbf{R}[k]
    \label{eq:psi-svd}
\end{equation}
Then we can write
\begin{equation}
    \mathbf{\Psi}[k]^{\mathrm{eff}} = \mathbf{\Psi}_1[k]^{\mathrm{eff}}
        \mathbf{S}[k]^{-1} \mathbf{\Psi}_2[k+1]^{\mathrm{eff}}
    \label{eq:psi-split}
\end{equation}
where
\begin{equation}
\begin{aligned}
    \mathbf{\Psi}_1[k]^{\mathrm{eff}} =&\ \mathbf{L}[k] \mathbf{S}[k] \\
    \mathbf{\Psi}_2[k+1]^{\mathrm{eff}} =&\ \mathbf{S}[k] \mathbf{R}[k]
\end{aligned}
\end{equation}
are the two new canonical centers at site \( k \) and \( k + 1 \). Once we have two canonical centers in the MPS, two partial DMRG sweeps, namely, a backward sweep starting from site \( k \) and a forward sweep starting from site \( k + 1 \), can be performed simultaneously by separate processors. The above approach can be invoked iteratively to generate \( P_{\mathrm{site}} \) canonical centers in the MPS, where \( P_{\mathrm{site}} \) is the total number of (groups of) processors at this level of parallelism.
Matrix \( \mathbf{S}[k]^{-1} \) (termed the connection matrix) is used after a round of forward and backward partial sweeps to merge the updated \( \mathbf{\Psi}^{\mathrm{new}}_1[k]^{\mathrm{eff}} \) and \( \mathbf{\Psi}^{\mathrm{new}}_2[k]^{\mathrm{eff}} \) to yield an approximation to the updated \( \mathbf{\Psi}^{\mathrm{new}}[k]^{\mathrm{eff}} \)
\begin{equation}
    \mathbf{\Psi}^{\mathrm{new}}[k]^{\mathrm{eff}} = \mathbf{\Psi}^{\mathrm{new}}_1[k]^{\mathrm{eff}}
        \mathbf{S}[k]^{-1} \mathbf{\Psi}^{\mathrm{new}}_2[k+1]^{\mathrm{eff}}
    \label{eq:wfn-merge}
\end{equation}
Merging  the two separately optimized portions of the MPS using this connection matrix does not change the MPS when the MPS has reached its variational optimum. 
The two partial sweeps over sites \( \cdots, k \) and \( k + 1, \cdots \) cannot update the MPS bond between the sites \( k \) and \( k + 1\). Therefore, a sweep iteration at the connection site is performed, where \autoref{eq:eff-eigsh} is solved for the merged wavefunction \( \mathbf{\Psi}^{\mathrm{new}}[k]^{\mathrm{eff}} \). The solution of \autoref{eq:eff-eigsh}, denoted as \( \mathbf{\Psi}^{\prime \mathrm{new}}[k]^{\mathrm{eff}} \), is then split according to \autoref{eq:psi-split} to generate the updated connection matrix \( \mathbf{S}^{\mathrm{new}}[k]^{-1} \).

In a typical \textit{ab initio} application, the amount of computation is not distributed homogeneously among different groups of sites (see \autoref{fig:nc}) because of the boundary effects of the MPS and the different block sizes from different truncations at different sites. In addition, the total number of sites available for this level of parallelism is limited. If the same number of sites is assigned to different processors, one observes a significant load imbalance, which negatively impacts the scalability.\cite{chen2021real}

To alleviate this problem, similar to the dynamic boundary strategy used in \lit{chen2021real}, we have added an additional step to dynamically determine the position of the canonical center (connection site) to improve load balancing. After all processors finish their partial sweeps, the total computational cost is measured for each processor for the partial sweep and all its sweep iterations. From this, it is possible to estimate whether changing the connection site from \( k \) to \( k + 2 \) (for example) reduces the cost discrepancy between the two processors connected at site \( k \). If this is the case, then the connection site is moved to \( k + 2 \), with the hope that this helps to reduce the degree of load imbalance during the next sweep.

We note that changing the position of the connection site between sweeps is not an operation with negligible cost, since not only the MPS tensors, but also the 
renormalized operators, need to be transformed (see \autoref{eq:op-trans}). Consequently, in our implementation, we have limited the distance between the old connection site and the new connection site to at most two sites. This ensures that the operation itself does not consume a significant amount of time. In practice, we can start the DMRG algorithm with an arbitrary set of connection sites. After several sweeps with dynamical adjustment of the connection sites, we observe that we can often achieve a stable set of connection sites and a well-balanced workload amongst the processors. The performance of the parallelism over sites with the dynamical adjustment of connection sites is discussed in \autoref{sec:dynamic}.

Comparing to the  strategy very recently introduced in \lit{chen2021real}, our approach does not directly reduce the waiting time of the current sweep; instead, the performance statistics of the current sweep are accumulated, to determine the position of the connection sites for the next sweep. In contrast, the approach introduced in \lit{chen2021real} completely removes the waiting time at each sweep, but introduces an extra projection error in the wavefunction initial guess (\autoref{eq:wfn-merge}) when the connection site is changed (which is larger for \textit{ab initio} systems compared to spin systems, according to \lit{chen2021real}). This extra error in the wavefunction transformation may increase the number of Davidson iterations. 

In order to improve single-node performance, we have also considered the fine-grained  strategies for shared memory parallelism.\cite{hager2004parallelization} Most of them can be easily implemented in an \textit{ab initio} DMRG code with minor modifications. These are now discussed.

\subsection{Shared memory parallelism over normal and complementary operators} \label{sec:para-shared-op}

The left-right decomposition of Hamiltonian (\autoref{eq:h-nc} and \autoref{eq:h-cn}) is a sum of products of normal and complementary operators. For the \textit{ab initio} sub-Hamiltonians, there are \( O(K^2/P_{\mathrm{hamil}} + K) \) terms in the summation. Therefore, for the matrix-vector multiplication
\begin{equation}
    |\Psi'[k]^{\mathrm{eff}}\rangle = \hat{H}[k]^{\mathrm{eff}}
        |\Psi[k]^{\mathrm{eff}}\rangle
\end{equation}
invoked during the Davidson procedure, we can divide the work among \( T_{\mathrm{op}} \) threads, namely
\begin{equation}
    \hat{H}[k]^{\mathrm{eff}} = \hat{H}[k]^{\mathrm{eff}}_{(1)} + 
        \hat{H}[k]^{\mathrm{eff}}_{(2)} + \cdots + \hat{H}[k]^{\mathrm{eff}}_{(T_{\mathrm{op}})}
\end{equation}
The partial contribution to \( |\Psi'[k]^{\mathrm{eff}}\rangle \) is computed on every thread \( t \) as
\begin{equation}
    |\Psi'[k]^{\mathrm{eff}}_{(t)}\rangle = \hat{H}[k]^{\mathrm{eff}}_{(t)}
        |\Psi[k]^{\mathrm{eff}}\rangle
    \label{eq:hpsi-thread}
\end{equation}
Finally, a reduction step is performed to obtain \( |\Psi'[k]^{\mathrm{eff}}\rangle \), as
\begin{equation}
    |\Psi'[k]^{\mathrm{eff}}\rangle =
        |\Psi'[k]^{\mathrm{eff}}_{(1)}\rangle + |\Psi'[k]^{\mathrm{eff}}_{(2)}\rangle
        +\cdots + |\Psi'[k]^{\mathrm{eff}}_{(T_{\mathrm{op}})}\rangle
\end{equation}
with a small additional computation cost of \( O(16M^2K \log T_{\mathrm{op}}) \) per sweep.

\subsection{Shared memory parallelism over symmetry sectors} \label{sec:para-sym}

In addition, every term in \autoref{eq:hpsi-thread} is implemented as a block-sparse matrix-matrix multiplication, which can be further decomposed into dense matrix-matrix multiplications over independent symmetry sectors. Instead of using nested threaded parallelism over normal and complementary operators and symmetry sectors, we can collapse the two thread parallelism levels to one level,\cite{brabec2020massively} to achieve a better load balance and reduce the overhead from creating threads.

\subsection{Shared memory parallelism within dense matrix multiplication} \label{sec:para-dense}

Thread-level parallelism in dense matrix multiplication can be easily introduced by using a threaded math library.\cite{hager2004parallelization} The effectiveness of this lowest level of parallelism is analyzed in \autoref{sec:dense}.

\subsection{Numerical Implementation}
\label{sec:impl}



We have used many low-level performance optimizations in the numerical implementation of the DMRG algorithm. We have reused many of the ideas in the \textsc{StackBlock} implementation,\cite{stackblock} which has been developed in our research group over many years.\cite{chan2002highly,chan2004algorithm,sharma2012spin} Specifically, we  store renormalized operators using stack memory to reduce memory fragmentation, we delay tensor contraction in the blocking step, and we use specialized routines for operations involving identity matrices or matrices with only a single non-zero element. In the new \textsc{Block2} code,\cite{block2} we also allow developers to switch off all these efficiency related optimizations, so that one can implement new methods without considering the effects of low-level optimizations.

In addition, we have introduced several improvements. 

(i)  We use symbolic algebra methods to construct and manipulate a symbolic MPO, where the elements of the MPO matrices are operator symbols (e.g. $1, a_2, \hat{R}^{L[\frac{1}{2}]}_3$ etc.; for a more complete list of symbols, see the left-right decomposition formulae in Appendix~\ref{app:qc-dmrg}). After the MPO construction, symbolic simplification can be used to generate optimal blocking formulae to multiply out the MPO, where
 related operator terms (e.g. $a_1 a_2 =-a_2 a_1$)  are merged,  and elements in the partially contracted MPO which do not contribute to the final Hamiltonian are eliminated. In addition, the symbolic algebra system supports the lazy contraction of renormalized operators during the DMRG algorithm.

(ii) Multiplication and tensor products of block-sparse matrices are handled lazily by building a list of the associated GEMM operations to be later dispatched using shared memory parallelism. 
The list of GEMMs is generated once for each site along the sweep. Also, because many matrices share the same block-sparse skeleton,  the list of block indices used during a contraction can be generated once for every combination of matrix skeletons. 
The delayed contraction, together with the use of stack memory, helps reduce the overhead of switching CPU contexts. Using this approach, we see a significant speed-up relative to the \textsc{StackBlock} code for small bond dimensions (\( M < 1000 \)).  At larger \( M \) the performance difference between the two implementations on a single node is small, since most time is spent in the math library. However, the reduced overhead in the \textsc{Block2} code is beneficial when scaling to thousands of CPU cores.

(iii) For large scale DMRG calculations, the total required storage and file input/output (IO) speed can be bottlenecks. 
To address this,  we directly dump and load stack memory chunks to avoid the CPU overhead of serialization and deserialization. In addition, we use floating point compression  during the disk IO. The compression  introduces a small user-defined per-number absolute error (\(10^{-14}\) in this work) in the stored renormalized operator matrix elements. We observed a 40\% reduction in required storage for the benzene system discussed below and no discernable effect on accuracy, while the total IO time was similar with or without  compression.


\section{Results} \label{sec:results}

As a first benchmark, we assess our parallel DMRG implementation in a ground-state energy calculation of benzene using a cc-pVDZ basis~\cite{dunning1989gaussian} with an orbital space comprising 108 orbitals and 30 electrons.\cite{eriksen2020ground} Although the benzene system is a closed shell system and thus does not showcase the strengths of the DMRG algorithm, it nonetheless serves as an example in the literature where a DMRG calculation with a large bond dimension and a relatively large number of orbitals has been recently reported.

For the benzene calculation, we use particle number, SU(2) (spin) and \( C_s \) point group symmetry to reduce the overall cost of the calculation. The same orbitals, integrals and orbital ordering as in \lit{eriksen2020ground} were used in this work. The DMRG correlation energy at \( M = 6000 \) (plus approximately 200 states to represent the low-weight quantum numbers) obtained in this work is \( -859.1 \ \mathrm{m}E_{H} \). Given the differences in implementation that gives rise to small differences in bond truncations across many sweeps, this is in excellent agreement with the DMRG correlation energy (\( -859.2 \ \mathrm{m}E_{H} \)) reported in \lit{eriksen2020ground} at \( M = 6000 \).

In addition, we demonstrate the performance of our DMRG implementation in a calculation on the FeMo cofactor system, using a model with 76 orbitals and 113 electrons in the active space recently proposed by Li et. al. in \lit{li2019femoco}. This is an example of a system with multiple transition metal centers where the strengths of the DMRG algorithm can in principle be demonstrated. We use the integral file provided in \lit{li2019femoco} without any further orbital reordering. The state with total spin \( S = 3/2 \) is targeted.

All calculations in this work use the two-site DMRG algorithm with perturbative noise.\cite{white2005density} Five sweeps were performed at each MPS bond dimension \( M \). To measure the wall time per sweep, we used the average wall time for the last four sweeps for each \( M \). For the benzene system, to alleviate the problem of losing quantum numbers, we kept at least one state for each quantum number after the normal decimation process.\cite{chan2004algorithm} This makes the bond dimension \( M \) in the calculation slightly larger than its specified target value. For example, when \( M \) 
was set to 6000, the observed actual \( M \) was typically about 6200.

We denote different parallelism schemes by a set of numbers \( P_{\mathrm{site}}, P_{\mathrm{hamil}}, T_{\mathrm{op}} \) and \( T_{\mathrm{dense}} \) indicating the number of groups of processors, processors, or threads used in the four levels of parallelism. Namely, \( P_{\mathrm{site}} \) denotes the parallelism over sites; \( P_{\mathrm{hamil}} \) denotes the distributed parallelism over sub-Hamiltonians; \( T_{\mathrm{op}} \) is for the joint shared memory parallelism over normal and complementary operators and symmetry sectors; and \( T_{\mathrm{dense}} \) is for the thread-level parallelism in the dense matrix multiplications. The total number of CPU cores for a specific parallelism scheme is given by \( N_{\mathrm{core}} = P_{\mathrm{site}} P_{\mathrm{hamil}} T_{\mathrm{op}} T_{\mathrm{dense}} \).

The calculations were executed on nodes with 28-core Intel Cascade Lake 8276 CPUs (2.20 GHz), made available via the Caltech high-performance computing facility. Each node has 56 CPU cores and 384 GB memory.

\subsection{Mixed NC/CN approach} \label{sec:mixed}

As discussed in \appref{app:qc-dmrg}, in conventional DMRG implementations, there are two possible ways to write the left-right decomposition of the \textit{ab initio} Hamiltonian at each site \( k \). The NC scheme corresponds to an MPO with tensor dimensions that increase from left to right, while the CN scheme corresponds to an MPO with tensor dimensions that decrease from left to right.\cite{chan2016matrix} Typically, efficient DMRG implementations use a mixed NC/CN approach,\cite{kurashige2009high} where the NC decomposition is used for sites \( k < K / 2 \) and the CN decomposition is used for sites \( k \ge K / 2 \), which gives a significantly smaller ``MPO" bond dimension. However, a transformation from the normal to complementary (two-index) operators is required near the middle site in this mixed NC/CN approach. The time complexity for this transformation is \( O(K^4M^2) \). Additionally, for parallelism over sub-Hamiltonians, since we use different processor assignments for the NC and CN schemes, an extra reduction step for all the two-index complementary operators is required. The communication cost is \( O(K^2M^2\log P_{\mathrm{hamil}}) \). The extra computation and communication cost means that the middle site of the sweep is significantly more expensive than the other sites. Consequently, for parallelism over sites, we consider only odd \( P_{\mathrm{site}} \) and use a non-uniform division of the sweep ranges, so that the high cost of computation at the middle site (included in the sweep range of processor group \( p = \lceil \frac{1}{2} P_{\mathrm{site}} \rceil \)) is amortized among all \( P_{\mathrm{site}} \) groups of processors. In this work, we tested \( P_{\mathrm{site}} = 1,3,5 \).

\begin{figure}[!htbp]
  \includegraphics[width=\columnwidth]{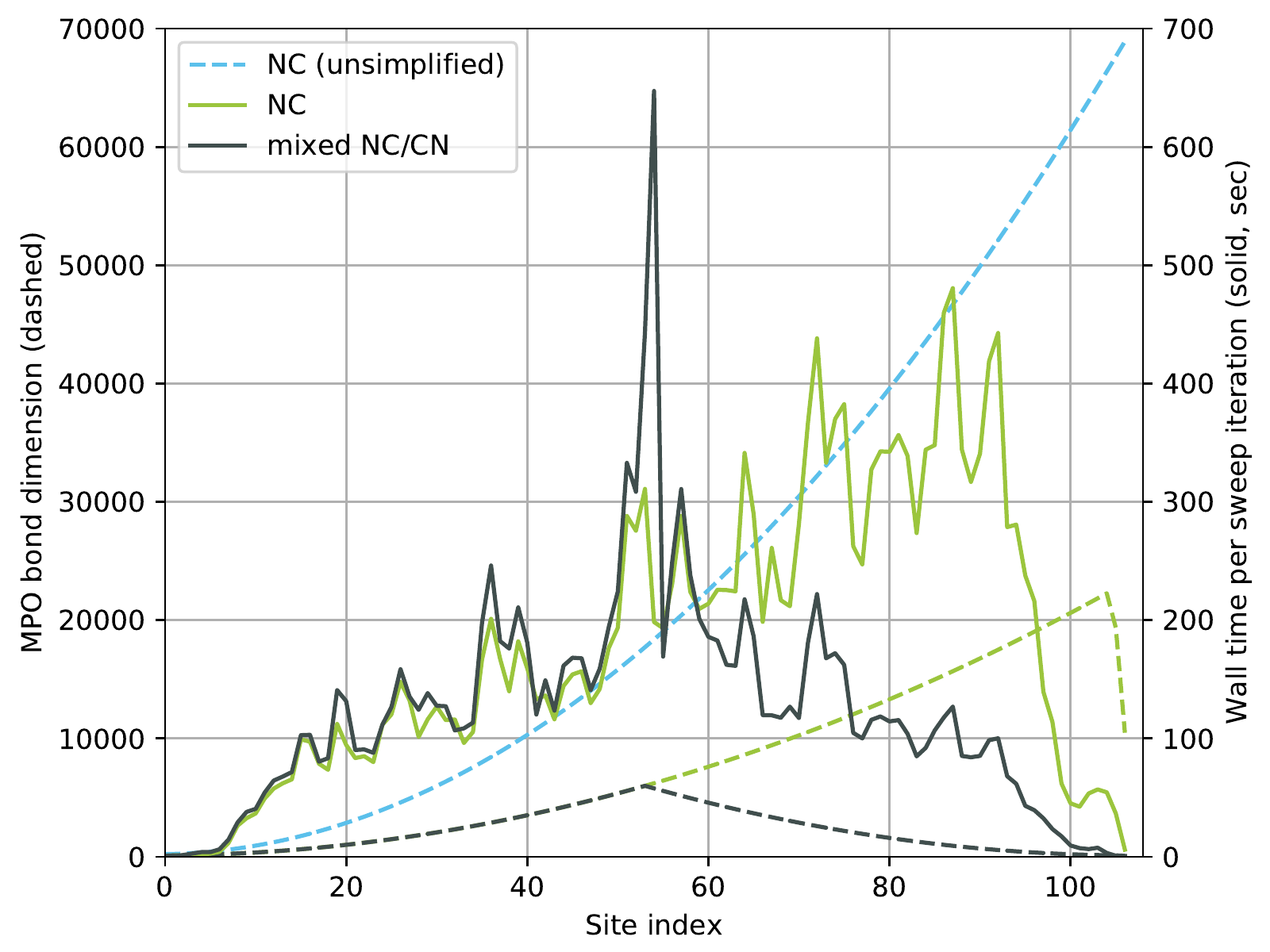}
  \caption{MPO tensor bond dimensions (dashed lines) and wall time cost (solid lines) at each site for the NC and mixed  NC/CN approaches for the benzene system. (Unsimplified refers to the bond dimension obtained without accounting for zero-integrals and symmetry conditions, given by $2+2K+6k^2$). The performance data is from a \( M = 4000 \) calculation with the parallelism scheme \( P_{\mathrm{site}} = 1, P_{\mathrm{hamil}} = 16, T_{\mathrm{op}} = 28\), and \( T_{\mathrm{dense}} = 1 \).}
  \label{fig:nc}
\end{figure}

In \autoref{fig:nc} we show the distribution of MPO bond dimensions (for each MPO tensor) and the corresponding wall time cost at each site, for the NC and mixed NC/CN partitions of the benzene system. For a spin-adapted DMRG algorithm, the bond dimension of the MPO tensor at site \( k \) is \( 2+2K + 6k^2 \) (blue dashed line), if the NC scheme is used without any optimization or additional simplifications. Using the symmetry conditions \autoref{eq:index-sym}, the bond dimension can be reduced to approximately \( 2+2K + 2k^2 \) (green dashed line), and the sudden decrease of the MPO bond dimension near the rightmost site in the figure is due to the removal of complementary operators with vanishing integrals. The mixed NC/CN approach gives a much better distribution of bond dimensions (black dashed line), with the maximal value \( D = 5996 \) appearing near the middle site of the test system. From \autoref{fig:nc} we can see that the time cost near the middle site of the mixed approach is approximately two times as large as that of the NC approach, but the mixed approach gives a much smaller total wall time per sweep (with \( M = 4000\) and \( P_{\mathrm{hamil}} = 16 \), \( t_{\mathrm{mixed}} = 13071 \mathrm{\ sec} \)) as compared to the NC approach (\( t_{\mathrm{nc}} = 19356 \mathrm{\ sec} \)), mainly due to the smaller MPO bond dimensions for \( k \ge K / 2 \). The speed-up \( t_{\mathrm{nc}}/t_{\mathrm{mixed}} \) is approximately 148\% based on the data in \autoref{fig:nc}. Due to this, subsequent calculations in this work all use the mixed approach.

\subsection{Parallelism within dense matrix multiplication} \label{sec:dense}

As discussed in earlier studies,\cite{hager2004parallelization}  using thread parallelism in the dense matrix multiplications is not very effective in DMRG, compared with the other parallel strategies. \autoref{tab:dense} shows that this is also true for our implementation of \textit{ab initio} DMRG. For \( M = 2500 \) and \( 3000 \), parallelism schemes with \( T_{\mathrm{dense}} = 4 \) are approximately 60\% to 70\% slower than the scheme with \( T_{\mathrm{dense}} = 1 \). Therefore, for production calculations in this work, this level of parallelism was not utilized (i.e., we used only \( T_{\mathrm{op}} = 28 \) and \( T_{\mathrm{dense}} = 1 \)).

\begin{table}[!htbp]
    \centering
    \caption{Wall time per sweep (in seconds) in the benzene calculation for MPS bond dimension \( M = 2500 \) and \( 3000 \) using parallelism schemes with different \( T_{\mathrm{dense}} \).}
    \label{tab:dense}
    \begin{tabular}{
        >{\centering\arraybackslash}p{0.8cm}
        >{\centering\arraybackslash}p{0.8cm}
        >{\centering\arraybackslash}p{0.8cm}
        >{\centering\arraybackslash}p{0.8cm}|
        >{\centering\arraybackslash}p{1.7cm}
        >{\centering\arraybackslash}p{2cm}}
        \hline\hline
    \multicolumn{4}{c|}{parallelism scheme} & \multicolumn{2}{c}{Wall time per sweep (sec)} \\ 
    \( P_{\mathrm{site}} \) & \( P_{\mathrm{hamil}} \) & \( T_{\mathrm{op}} \) & \( T_{\mathrm{dense}} \) & \( M = 2500 \) & \( M = 3000 \) \\
    \hline
    5& 14 & 28 & 1 & 893 & 1291 \\
    5& 14 & 7 & 4 & 1521 & 2106 \\
    5& 7 & 14 & 4 & 1493 & 2079 \\
    \hline\hline
    \end{tabular}
\end{table}

\subsection{Parallelism over sites} \label{sec:dynamic}

\begin{figure}[!htbp]
  \includegraphics[width=\columnwidth]{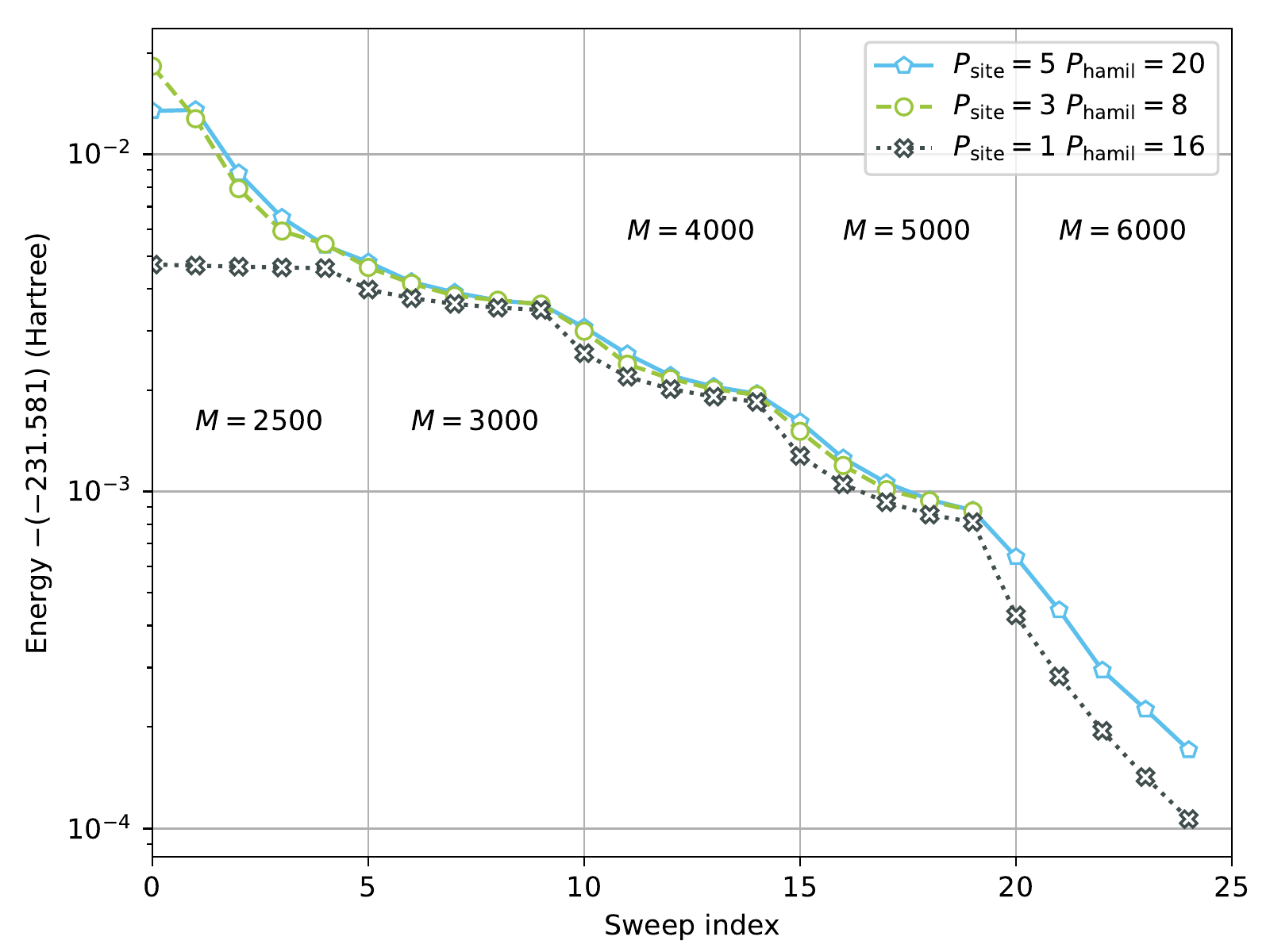}
  \caption{Sweep energies for parallelism schemes with different \( P_{\mathrm{site}} \) and different MPS bond dimensions \( M \) for the benzene system. For each \( M \), five sweeps are performed. }
  \label{fig:ener}
\end{figure}

It is sometimes argued that when parallelism over sites is used, the convergence of the DMRG energy as a function of the number of sweeps is slower than that of the standard DMRG approach, if the same sweep schedule is used.\cite{stoudenmire2013real} In \autoref{fig:ener} we compare parallelism schemes with different \( P_{\mathrm{site}} \) for MPS bond dimensions up to \( M = 6000 \) (data for \( M = 6000 \) with \( P_{\mathrm{site}} = 3 \) and \( P_{\mathrm{hamil}} = 8 \) could not be obtained due to memory constraints) for the benzene system. We can see that in our test system,  convergence is only slightly affected by increasing \( P_{\mathrm{site}} \) from 1 to 5. When five sweeps were performed for each \( M \), the energy obtained from the last sweep for each \( M \) was almost the same with different \( P_{\mathrm{site}} \), up to \( M = 5000 \). Although we started the calculation from the same initial MPS (with a single canonical center) for different \( P_{\mathrm{site}} \), for \( P_{\mathrm{site}} = 3\) and 5 the initial MPS is re-canonicalized to introduce extra canonical centers. During this canonicalization step some low-weight single-state quantum numbers were discarded, which makes the \( P_{\mathrm{site}} = 3\) and 5 DMRG energy at \( M = 2500 \) (artificially) higher in \autoref{fig:ener}.

\begin{figure}[!htbp]
  \includegraphics[width=\columnwidth]{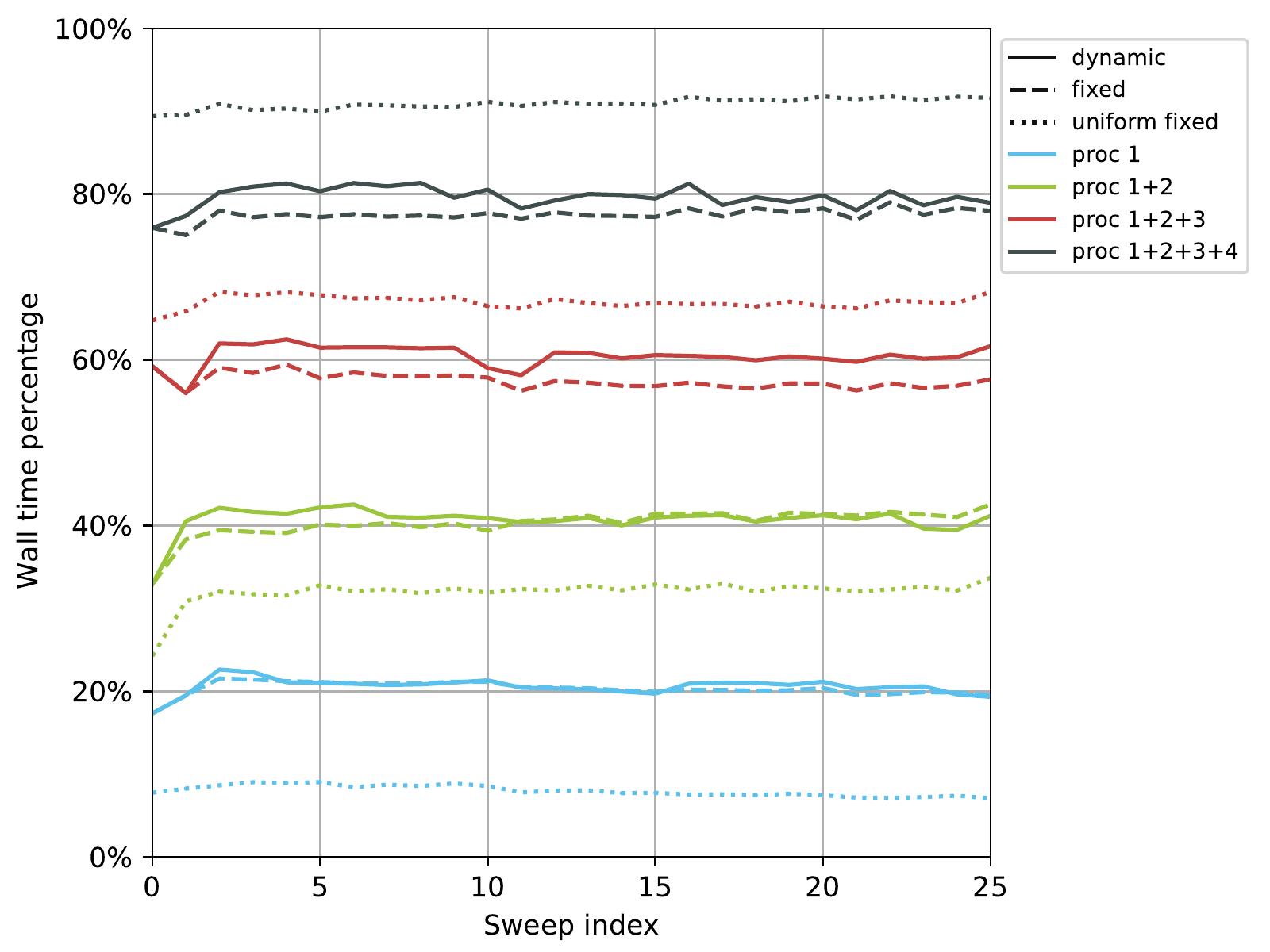}
  \caption{Estimated wall time per processor for parallelism over sites (as a percentage of the sum of wall times for all processors), when the connection centers are dynamically adjusted (solid lines), fixed (dashed lines), uniformly distributed (dotted lines). The performance data is from the \( P_{\mathrm{site}} = 5\) and \( P_{\mathrm{hamil}} = 20 \) benzene calculation with the MPS bond dimension increasing from \( M = 2500 \) to \( M = 6000 \). }
  \label{fig:dyn}
\end{figure}

To examine the effect of the dynamical connection sites, we have compared the estimated performance using dynamical, fixed and uniform connection sites in \autoref{fig:dyn}. For our test benzene system with 108 sites and \( P_{\mathrm{site}} = 5 \), four connection sites are required. From the dotted lines in \autoref{fig:dyn}, we  see  selecting connection sites based on a uniform division of sweep ranges (namely, \( K_{\mathrm{conn}} = \{ 21, 43, 64, 86 \} \)) gives a large load imbalance among the five processors. At the last sweep, the longest processor task consumed 388\% more time than the shortest processor task, which can be mainly attributed to the highly non-uniform distribution of computational effort among sites (see solid black line in \autoref{fig:nc}). In this work, we found that \( K_{\mathrm{conn}} = \{ 33, 49, 57, 73 \} \) (obtained from using dynamical connection sites for small bond dimensions) gave much better performance. This corresponds to the dashed lines in \autoref{fig:dyn}. With this fixed set of connection sites, the longest task consumed 53\% more time than the shortest task. If we allow the set of connection sites to be dynamically adjusted between the sweeps, we end up with a slightly altered set of connection sites \( K_{\mathrm{conn}} = \{ 33, 48, 59, 74 \} \). Using this, in the last sweep the longest task then consumed only 26\% more time than the shortest task.

\subsection{Parallel Scaling}

\begin{figure}[!htbp]
  \includegraphics[width=\columnwidth]{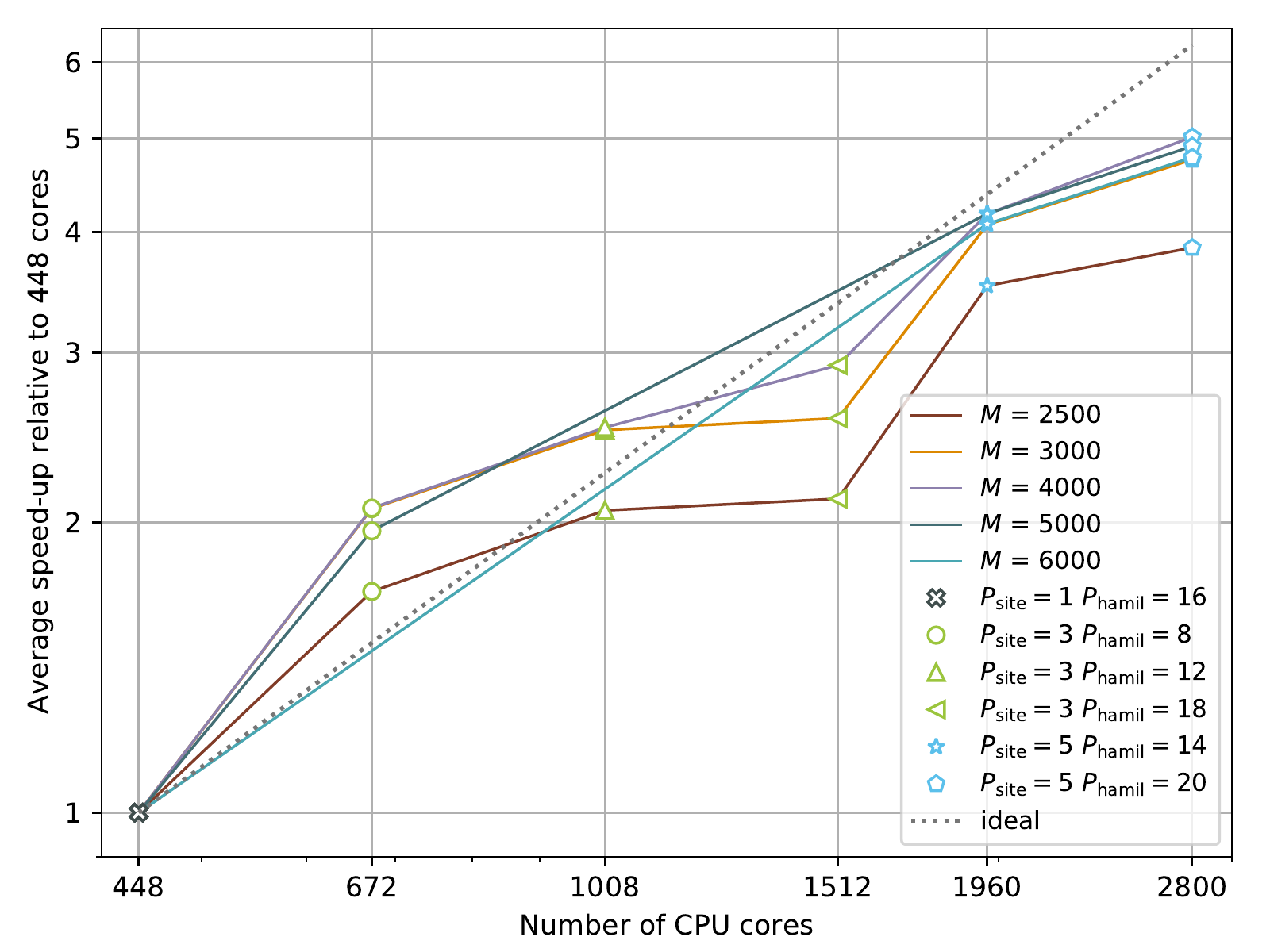}
  \caption{Speed-up of average wall time per sweep relative to the \( N_{\mathrm{core}} = 448 \) case for different MPS bond dimensions using parallelism schemes with different \( P_{\mathrm{site}} \) and \( P_{\mathrm{hamil}} \). \( T_{\mathrm{op}} = 28\) and \( T_{\mathrm{dense}} = 1 \) are used for all parallelism schemes.}
  \label{fig:perf}
\end{figure}

In \autoref{tab:scaling} we list the average wall time per sweep with MPS bond dimensions from \( M = 2500 \) to \( M = 6000 \) for the benzene system, when parallelism schemes with different \( P_{\mathrm{site}} \) and \( P_{\mathrm{hamil}} \) are used. The speed-up relative to the \( P_{\mathrm{site}} = 1 \) and \( P_{\mathrm{hamil}} = 16 \) (\( N_{\mathrm{core}} = 448 \)) case is plotted in \autoref{fig:perf}.

\begin{table*}[!htbp]
    \centering
    \caption{Average wall time per sweep (in seconds) of the benzene calculation for different MPS bond dimensions using parallelism schemes with different \( P_{\mathrm{site}} \) and \( P_{\mathrm{hamil}} \). \( T_{\mathrm{op}} = 28\) and \( T_{\mathrm{dense}} = 1 \) were used for all parallelism schemes.}
    \label{tab:scaling}
    \begin{tabular}{
        >{\centering\arraybackslash}p{1.2cm}
        >{\centering\arraybackslash}p{1.2cm}|
        >{\centering\arraybackslash}p{1.2cm}|
        >{\centering\arraybackslash}p{1.5cm}
        >{\centering\arraybackslash}p{1.5cm}
        >{\centering\arraybackslash}p{1.5cm}
        >{\centering\arraybackslash}p{1.5cm}
        >{\centering\arraybackslash}p{1.5cm}
        }
        \hline\hline
    \multicolumn{2}{c|}{parallelism scheme} & 
    \multirow{2}{*}{\( N_{\mathrm{core}} \)} &
    \multicolumn{5}{c}{Average wall time per sweep (sec)} \\ 
    \( P_{\mathrm{site}} \) & \( P_{\mathrm{hamil}} \) &
    & \( M = 2500 \) & \( M = 3000 \) & \( M = 4000 \)
    & \( M = 5000 \) & \( M = 6000 \) \\
    \hline
    1 & 16 & 448 & 3145 & 5253 & 12740 & 22212 & 35451 \\
    \rule{0pt}{3.5ex} \multirow{3}{*}{3} & 8 & 672 & 1855 & 2542 & 6158 & 11335 & \\
     & 12 & 1008 & 1529 & 2107 & 5079 & & \\
     & 18 & 1512 & 1487 & 2049 & 4379 & & \\
    \rule{0pt}{3.5ex} \multirow{2}{*}{5} & 14 & 1960 & 894 & 1291 & 3051 & 5317 & 8696 \\
     & 20 & 2800 & 816 & 1105 & 2539 & 4526 & 7419 \\
    \hline\hline
    \end{tabular}
\end{table*}

\begin{table*}[!htbp]
    \centering
    \caption{Average wall time per sweep (in seconds) of the FeMo cofactor calculation for different MPS bond dimensions using parallelism schemes with different \( P_{\mathrm{site}} \) and \( P_{\mathrm{hamil}} \). \( T_{\mathrm{op}} = 28\) and \( T_{\mathrm{dense}} = 1 \) were used for all parallelism schemes.}
    \label{tab:scaling-femoco}
    \begin{tabular}{
        >{\centering\arraybackslash}p{1.2cm}
        >{\centering\arraybackslash}p{1.2cm}|
        >{\centering\arraybackslash}p{1.2cm}|
        >{\centering\arraybackslash}p{1.5cm}
        >{\centering\arraybackslash}p{1.5cm}
        >{\centering\arraybackslash}p{1.5cm}
        }
        \hline\hline
    \multicolumn{2}{c|}{parallelism scheme} & 
    \multirow{2}{*}{\( N_{\mathrm{core}} \)} &
    \multicolumn{3}{c}{Average wall time per sweep (sec)} \\ 
    \( P_{\mathrm{site}} \) & \( P_{\mathrm{hamil}} \) &
    & \( M = 2000 \) & \( M = 2500 \) & \( M = 3000 \) \\
    \hline
    1 & 16 & 448 & 10596 & 19464 & 50677 \\
    3 & 8  & 672 & 6380 & 12191 & 31496 \\
    5 & 16 & 2240 & 3262 & 5156 & 12499 \\
    \hline\hline
    \end{tabular}
\end{table*}

In \autoref{fig:perf} we see that, when \( P_{\mathrm{site}} = 3 \), increasing \( P_{\mathrm{hamil}} \) from 12 to 18 only reduces the wall time slightly, while nearly ideal speed-up is observed for different \( (P_{\mathrm{site}}, P_{\mathrm{hamil}}) \) when increasing from (1, 16) to (3, 12) and from (3, 12) to (5, 14). This illustrates that a combination of different DMRG parallelism strategies is essential to achieve good scaling across thousands of CPU cores. The better-than-ideal speed-up for \( (P_{\mathrm{site}}, P_{\mathrm{hamil}}) \) when increasing from (1, 16) to (3, 8) is also due to the change of parallelism strategy. Ideally, the \( P_{\mathrm{site}} = P_{\mathrm{hamil}} = 1 \) case should be used as the reference point for computing the speed-up. However, this is not feasible in our test system due to the large maximum MPO bond dimension \( D = 5996 \) and when using a large MPS bond dimension \( M = 6000 \). Here, we need \( P_{\mathrm{hamil}} \ge 10 \) to ensure that the memory cost per node is less than 384 GB.
This is an important reason to use larger \( P_{\mathrm{hamil}} \) rather than \( P_{\mathrm{site}} \) in certain systems, since increasing \( P_{\mathrm{site}} \) does not reduce the memory cost per processor group. Finally, we note that the speed-up for \( M=2500 \) appears to be significantly less than the other cases with the larger \( M \). This is likely related to the fact that for the \( P_{\mathrm{site}} = 3 \) and \( P_{\mathrm{site}} = 5 \) cases, an initial \( M = 2500 \) MPS with an artificially higher energy
was used (see \autoref{fig:ener}).

For the largest calculation considered in this work with \( P_{\mathrm{site}} = 5, P_{\mathrm{hamil}} = 20 \) and \( M = 6000 \) for the benzene system, the average communication and idle time among the \( P_{\mathrm{hamil}} \) processors constituted approximately 15\% of the total wall time for each group of \( P_{\mathrm{hamil}} \) processors and the average idle time among the \( P_{\mathrm{site}} \) groups of processors was approximately 10\% of the total wall time. The Davidson step (including communication) constituted 60\% to 70\% of the total wall time for each processor. Reading/writing disk files cost approximately 5\% of the total wall time.

In \autoref{tab:scaling-femoco} we show that a similar scaling can be observed for the FeMo cofactor system. When increasing \( (P_{\mathrm{site}}, P_{\mathrm{hamil}}) \) from \( (1, 16) \) to \( (5, 16) \), for a sufficiently large MPS bond dimension (\( M = 3000 \)) we obtain a speed-up of 4.05, which is close to the ideal speed-up (5). Note that the worse-than-cubic scaling with respect to \( M \) for the \( M = 2500 \) and \( M = 3000 \) cases shown in \autoref{tab:scaling-femoco} is mainly due to the difference in the Davidson convergence criteria used for different \( M \).

Finally, we note that using a similar number of CPU cores (\( N_{\mathrm{core}} \approx 2000 \)) and the same bond dimension (\( M = 3000 \)), the FeMo cofactor system wall time per sweep was 10 times larger than that for the benzene system. Although a slightly higher point group symmetry (\( C_s \)) was used in the benzene calculations, the main reason for this difference appears to be the strength of correlation in the FeMo cofactor. This leads to more Davidson iterations at each site, associated with the increased DMRG truncation error (\( 7\times 10^{-4} \) for the FeMo cofactor versus \( 1\times 10^{-5} \) for benzene) which means that the quality of the initial Davidson guess wavefunction is  poorer, when moving from site to site in the sweep. 

\section{Conclusions} \label{sec:conclude}

In this work, we introduced a modification of the conventional strategy for distributed memory parallelism in \textit{ab initio} DMRG algorithms that reduces the computation to the manipulation of independent sub-Hamiltonians, together with a small wavefunction communication step. 
This formulation thus combines the conceptual advantages of the sum of sub-Hamiltonians approach introduced in earlier work, with the greater parallelizability and lower prefactor of the conventional distributed memory DMRG algorithm. In addition, we carried out a comprehensive examination and implementation of four other sources of parallelism in DMRG, introducing techniques for load balancing via dynamic connection sites in site-based parallelism, and collapsing tasks to maximize thread efficiency in the shared memory parallelism. 
Finally, we showed that the combination of different DMRG parallelism strategies using both distributed and shared memory models was essential to achieve near-ideal speed-ups for a benchmark calculation with 108 orbitals and a DMRG bond dimension of \( M = 6000 \), scaling from 448 to 2800 CPU cores.

The DMRG implementation in the \textsc{Block2} code used in this work is open-source and can be freely obtained.\cite{block2} In addition to the ground state DMRG described here, it also supports several other MPS algorithms for \textit{ab initio} systems, including finite-temperature DMRG, imaginary and real time evolution, and reduced density matrix and transition density matrix evaluation.  Applications using these algorithms will be explored in future work.


\begin{acknowledgments}
This work was supported by the US National Science Foundation (NSF) via grant CHE-2102505. HZ thanks Seunghoon Lee for providing the integrals and reference DMRG outputs for the benzene system, and Henrik R.~Larsson, Zhi-Hao Cui and Tianyu Zhu for helpful discussions. The computations presented in this work were conducted on the Caltech High Performance Cluster, partially supported by a grant from the Gordon and Betty Moore Foundation.
\end{acknowledgments}

\section*{Data Availability}
The performance data presented in this work can be reproduced using the \textsc{Block2} code\cite{block2} and the integral files\footnote{The integral file for the benzene calculation can be found in \url{https://github.com/seunghoonlee89/SI-benzene-paper-DMRG}. The integral file for the FeMo cofactor calculation can be found in \url{https://github.com/zhendongli2008/Active-space-model-for-FeMoco}} provided in \lit{eriksen2020ground} and \lit{li2019femoco}.

\appendix
\section{The serial DMRG algorithm}
\label{app:dmrg}

To establish notation for the DMRG algorithm, consider a quantum lattice system with \( K \) sites. Each site is associated with a Hilbert space spanned by a site-basis \( \{ |n_k\rangle \} \). A complete basis of the system Hilbert space can be defined as the tensor product of \( K \) site-bases
\begin{equation}
    \{ |n_1\ n_2\ \cdots \ n_K \rangle \}
        = \{ |n_1\rangle \otimes |n_2\rangle \otimes \cdots \otimes |n_K\rangle \}
\end{equation}
The goal of the DMRG algorithm is to optimize a variational wavefunction in this Hilbert space, whose amplitudes can be written as a product of matrices
\begin{equation}
    |\Psi\rangle = \sum_{\{n\}} \mathbf{A}[1]^{n_1} \mathbf{A}[2]^{n_2} \cdots       
        \mathbf{A}[K]^{n_K} |n_1\ n_2\ \cdots \ n_K \rangle
\end{equation}
where each \( \mathbf{A}[k]^{n_k}\ (k = 2,\cdots, K -1) \) is an \( M \times M \) matrix, and the leftmost and rightmost matrices are \( 1 \times M \) and \( M \times 1 \) vectors, respectively. The dimension \( M \) is known as the bond-dimension of the MPS \( |\Psi\rangle \).

Within the MPS ansatz,  variational minimization of the energy, 
formally written as
\begin{equation}
    E_0 = \min_{|\Psi\rangle}
        \frac{ \langle \Psi|\hat{H} | \Psi\rangle }{ \langle \Psi|\Psi\rangle }
\end{equation}
where \( \hat{H} \) is the system Hamiltonian and \( E_0 \) is the ground-state energy,
can be performed iteratively by  optimizing the parameters of
a single matrix at a time in the MPS, 
while the parameters in the remaining matrices are kept constant. This corresponds to the 1-site DMRG algorithm. A common variant, designed to improve the ability to escape local minima,  optimizes a single larger matrix $A[k]^{n_k n_{k+1}}$ that describes the variational space of 2-sites at a time. This formally takes one outside of the single-site MPS variational space and thus the solution must be decimated back to the standard MPS form. This corresponds to the 2-site DMRG algorithm. The same idea can be generalized to $d$ sites,
but in this work we mainly consider the \( d = 2 \) case.

\begin{figure}[!htbp]
  \includegraphics[width=\columnwidth]{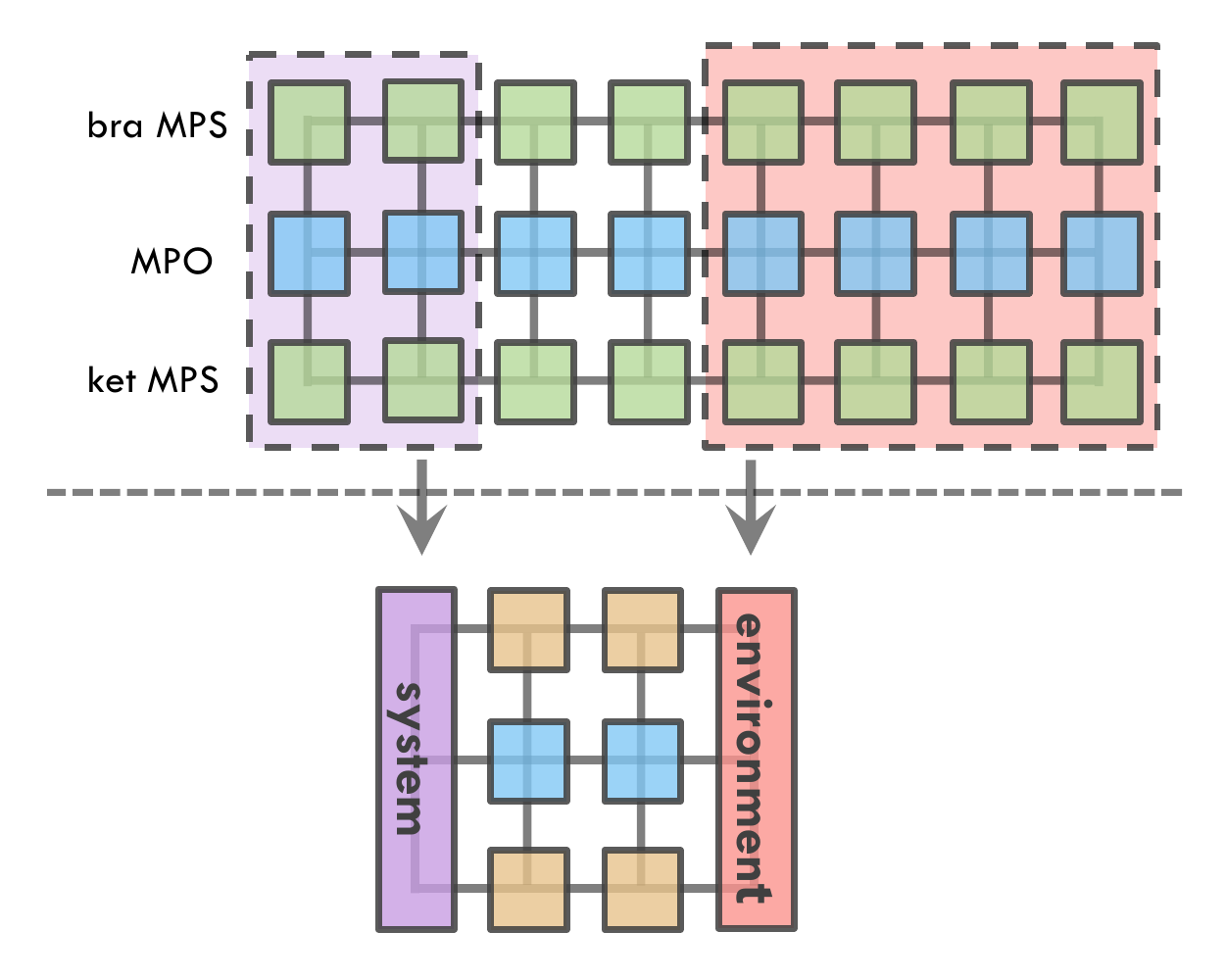}
  \caption{The left block (system), right block (environment) and 
  the individual sites being optimized in a given sweep iteration of the 2-site DMRG algorithm.}
  \label{fig:dmrg1}
\end{figure}

The iterative process in a serial DMRG algorithm is structured as a series of \textit{sweeps} along a fixed one-dimensional ordering of the \( K \) sites. Each sweep alternates between the forward and backward directions, consisting of \( K + 1 - p \) \textit{sweep iterations}. In the \( k \)-th \( (k=1,\cdots, K+1-p) \) sweep iteration of a forward sweep, the parameters in the
current matrix being optimized (associated with 
\( d \) adjacent sites,  \( \mathbf{A}[k]^{n_k \ldots n_{k+d-1}} \)) 
are updated, while in a backward sweep the matrices are updated in  reverse order. The lattice can then be conveniently divided into \( 2 + d \) \textit{blocks} (or sets of sites) \( \{ L_{k-1},S_k,\cdots,S_{k+d-1},R_{k+d} \} \) in the \( k \)-th sweep iteration (of a forward sweep, for example): a left block (or the \textit{system}) \( L_{k-1} \) for sites \( 1, \cdots, k - 1 \);  the individual sites whose matrices are being optimized \( S_k \) ... \( S_{k+d-1} \); and the right block (or the \textit{environment}) \( R_{k+d} \) for sites \( k + d, \cdots, K \) (see \autoref{fig:dmrg1}).

In each sweep iteration, we consider a left-right decomposition of the system Hamiltonian as the sum of tensor products of operators defined in blocks \( L_k \) and \( R_{k+1} \)
\begin{equation}
    \hat{H}[k] = \hat{H}^{L_k} \otimes \hat{1}^{R_{k+1}} + \hat{1}^{L_k} \otimes \hat{H}^{R_{k+1}} + \sum_{i} \hat{h}_i^{L_k} \hat{h}_i^{R_{k+1}}
    \label{eq:lr-decomp}
\end{equation}
where a bipartition of the lattice \( \{ L_k, R_{k+1} \} \) has been used. A convenient way to construct this left-right decomposition for any \( k \) is to first write the system Hamiltonian in a so-called MPO form
\begin{multline}
    \hat{H} = \sum_{\{n,n'\}} \mathbf{W}[1]^{n_1n'_1} \mathbf{W}[2]^{n_2n'_2} \cdots
        \mathbf{W}[K]^{n_Kn'_K} \\
         \times |n_1\ n_2\ \cdots \ n_K \rangle \langle n'_1\ n'_2\ \cdots \ n'_K |
    \label{eq:h-mpo}
\end{multline}
where each \( \mathbf{W}[k]^{n_kn'_k}\ (k = 2,\cdots, K -1) \) is a \( D \times D' \) matrix, and the leftmost and rightmost matrices are \( 1 \times D' \) and \( D \times 1 \) vectors, respectively. The maximal dimension \( D \) among these matrices will be called the bond-dimension of the MPO.

The left-right decomposition of the MPS can be defined as (in 2-site DMRG, for example)
\begin{multline}
    |\Psi[k]\rangle = \sum_{\alpha_{k-1} \alpha_k \alpha_{k+1}, n_k n_{k+1}}
        A[k]^{n_k}_{\alpha_{k-1}\alpha_k}
        A[k+1]^{n_{k+1}}_{\alpha_{k}\alpha_{k+1}} \\
        \times |\alpha_{k-1}^L\rangle \otimes
        |n_k\ n_{k+1}\rangle \otimes |\alpha_{k+1}^R\rangle
    \label{eq:eff-mps}
\end{multline}
where the left and right renormalized basis vectors are
\begin{equation}
\begin{aligned}
    |\alpha_{k}^L\rangle =&\ \sum_{\{n_1\cdots n_k\}}
        \bigg[ \mathbf{A}[1]^{n_1} \cdots
        \mathbf{A}[k]^{n_k} \bigg]_{\alpha_k} |n_1\ \cdots \ n_k \rangle \\
    |\alpha_{k}^R\rangle =&\ \sum_{\{n_{k+1}\cdots n_K\}}
        \bigg[ \mathbf{A}[k+1]^{n_{k+1}} \cdots
        \mathbf{A}[K]^{n_K} \bigg]_{\alpha_k} \\
        &\ \quad \quad \quad \quad \quad \quad \quad \quad
        \times |n_{k+1}\ \cdots \ n_K \rangle
\end{aligned}
\end{equation}

Using the MPO form, the decomposition \autoref{eq:lr-decomp} can be constructed as
\begin{equation}
    \hat{H}[k] = \sum_{\beta_k} \hat{H}[k]^L_{\beta_k} \otimes \hat{H}[k]^R_{\beta_k}
    \label{eq:lr-decomp-mpo}
\end{equation}
where
\begin{equation}
\begin{aligned}
    \hat{H}[k]^L_{\beta_k} =&\
        \sum_{\{n_1\cdots n_k,n'_1\cdots n'_k\}} \bigg[ \mathbf{W}[1]^{n_1n'_1} \cdots
        \mathbf{W}[k]^{n_kn'_k} \bigg]_{\beta_k} \\
        &\ \quad \quad \quad \quad \quad \quad 
            \times |n_1\cdots n_k \rangle \langle n'_1\cdots n'_k| \\
    \hat{H}[k]^R_{\beta_k} =&\
        \sum_{\{n_{k+1}\cdots n_K,n'_{k+1}\cdots n'_K\}} \bigg[ \mathbf{W}[k+1]^{n_{k+1}n'_{k+1}} \cdots \\
        &\ \times \mathbf{W}[K]^{n_Kn'_K} \bigg]_{\beta_k}
        |n_{k+1}\cdots n_K \rangle \langle n'_{k+1}\cdots n'_K|
    \label{eq:h-eff-lr}
\end{aligned}
\end{equation}


\begin{figure}[!htbp]
  \includegraphics[width=\columnwidth]{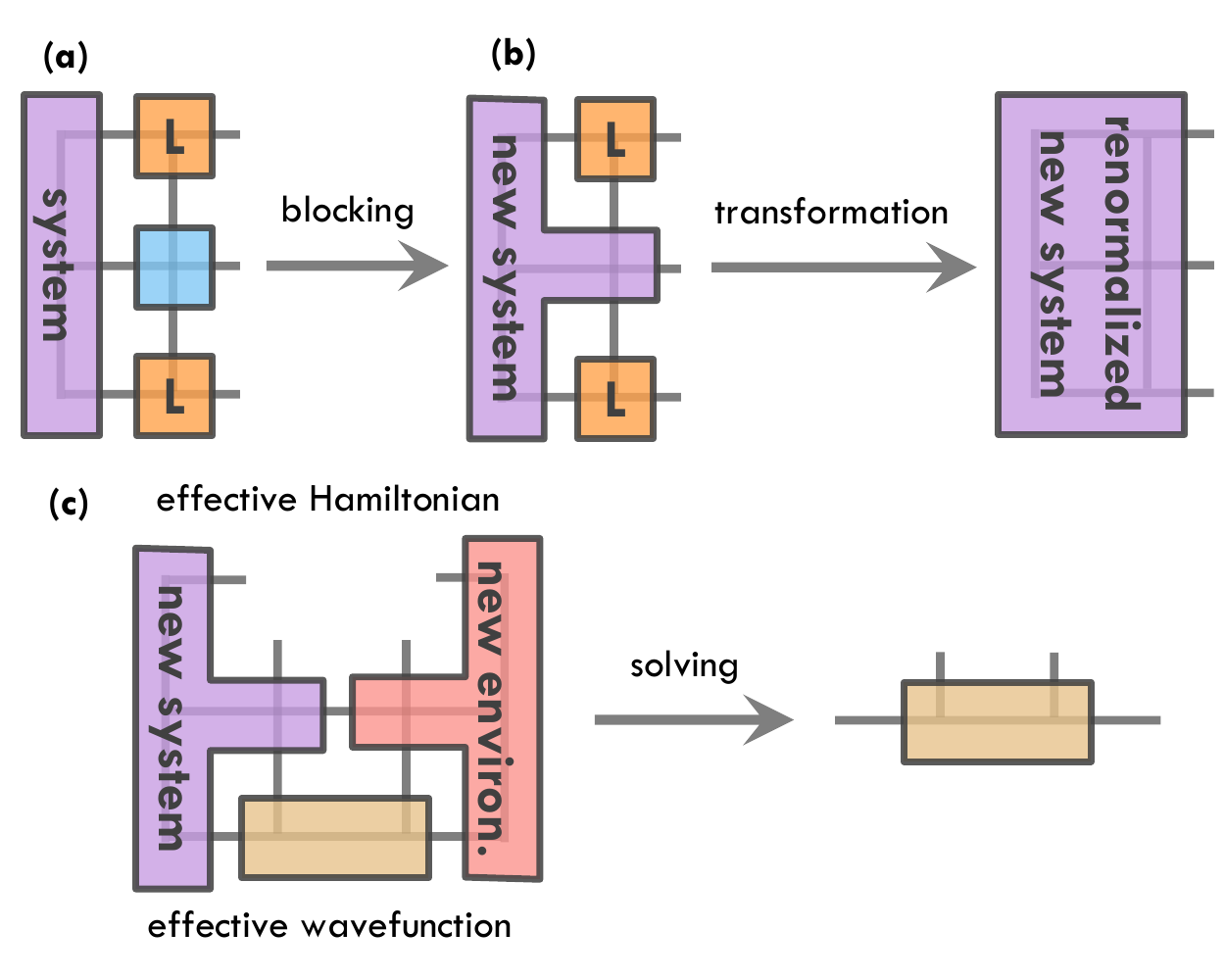}
  \caption{The (a) blocking, (b) transformation, and (c) solving steps in each sweep iteration of the 2-site DMRG algorithm.\cite{chan2016matrix}}
  \label{fig:dmrg2}
\end{figure}

Each sweep iteration of the 2-site DMRG algorithm is divided into three main steps (see \autoref{fig:dmrg2}):\cite{chan2002highly}

(i) blocking, where we compute the matrix representation of \( \hat{H}[k]_{\beta_k}^{L} \) and \( \hat{H}[k]_{\beta_k}^{R} \) \autoref{eq:h-eff-lr}) in bases \( |\alpha_{k-1}^L\ n_k\rangle \) and \( |n_{k+1}\ \alpha_{k+1}^R\rangle \) from the renormalized operators represented in bases \( |\alpha_{k-1}^L\rangle \) and \( |\alpha_{k+1}^R\rangle \), respectively
\begin{equation}
\begin{aligned}
    &\ \langle \alpha_{k-1}^{L}\ n_k | \hat{H}[k]_{\beta_k}^{L} |\alpha_{k-1}^{\prime L}\ n'_k\rangle \\
    =&\ \sum_{\beta_{k-1}} W[k]^{n_k n'_k}_{\beta_{k-1}\beta_k}
        \langle \alpha_{k-1}^L | \hat{H}[k-1]_{\beta_{k-1}}^{L} |\alpha_{k-1}^{\prime L} \rangle \\
    &\ \langle n_{k+1}\ \alpha_{k+1}^R | \hat{H}[k]_{\beta_k}^{R}
        |n'_{k+1}\ \alpha_{k+1}^{\prime R} \rangle \\
    =&\ \sum_{\beta_{k+1}} W[k+1]^{n_{k+1} n'_{k+1}}_{\beta_k\beta_{k+1}}
        \langle \alpha_{k+1}^R | \hat{H}[k+1]_{\beta_{k+1}}^{R} |\alpha_{k+1}^{\prime R} \rangle
    \label{eq:blocked-ops}
\end{aligned}
\end{equation}

(ii) solving, where we update the wavefunction in the renormalized basis
    \( |\alpha_{k-1}^L\ n_k\rangle \otimes |n_{k+1}\ \alpha_{k+1}^R\rangle \) (\autoref{eq:eff-mps}), given by
\begin{equation}
    \Psi[k]_{\alpha_{k-1}n_k,n_{k+1}\alpha_{k+1}}^{\mathrm{eff}} = \sum_{\alpha_k}
        A[k]^{n_k}_{\alpha_{k-1}\alpha_k}
        A[k+1]^{n_{k+1}}_{\alpha_{k}\alpha_{k+1}}
    \label{eq:eff-wfn}
\end{equation}
by solving the eigenvalue problem
\begin{equation}
    \mathbf{H}[k]^{\mathrm{eff}} \mathbf{\Psi}[k]^{\mathrm{eff}}
        = E[k] \mathbf{\Psi}[k]^{\mathrm{eff}}
    \label{eq:eff-eigsh}
\end{equation}
where the matrix elements of the effective Hamiltonian \( \mathbf{H}[k]^{\mathrm{eff}} \) are given by
\begin{multline}
    H[k]^{\mathrm{eff}}_{\alpha_{k-1}n_k,n_{k+1}\alpha_{k+1};
        n'_{k+1}\alpha'_{k+1},\alpha'_{k-1}n'_k} \\
        = \sum_{\beta_k} \langle\alpha_{k-1}^{L}\ n_k | \hat{H}[k]_{\beta_k}^{L} |\alpha_{k-1}^{\prime L}\ n'_k\rangle \\
        \times \langle n_{k+1}\ \alpha_{k+1}^R | \hat{H}[k]_{\beta_k}^{R}
        |n'_{k+1}\ \alpha_{k+1}^{\prime R} \rangle
\end{multline}
Since the Hamiltonian is sparse, the eigenvalue problem is normally solved using an iterative method such as the Davidson algorithm.\cite{davidson1975iterative}

(iii) decimation and transformation. Once the optimized wavefunction \( \mathbf{\Psi}[k]^{\mathrm{eff}} \) is determined, the new \( \mathbf{A}[k] \) and \( \mathbf{A}[k+1] \) can be found by decomposing the wavefunction using the density matrix, or via a singular value decomposition (SVD). After the decomposition, the matrix dimensions of \( \mathbf{A}[k] \) and \( \mathbf{A}[k+1] \) are truncated to bond dimension \( M \) by discarding small singular values or eigenvalues.
The truncated \( \mathbf{A}[k] \) and \( \mathbf{A}[k+1] \) are then used to construct new renormalized bases \( |\alpha_{k}^L\rangle \) and \( |\alpha_{k}^R\rangle \), in a forward and backward sweep iteration, respectively, as
\begin{equation}
\begin{aligned}
    |\alpha_{k}^L\rangle =&\ \sum_{\alpha_{k-1}}
        A[k]^{n_k}_{\alpha_{k-1}\alpha_k} |\alpha_{k-1}^L\ n_k \rangle \\
    |\alpha_{k}^R\rangle =&\ \sum_{\alpha_{k+1}}
        A[k+1]^{n_{k+1}}_{\alpha_k\alpha_{k+1}} |n_{k+1}\ \alpha_{k+1}^R \rangle
\end{aligned}
\end{equation}
The operators formed in the blocking step (\autoref{eq:blocked-ops}) are also transformed to the new renormalized basis
\begin{equation}
\begin{aligned}
    &\ \langle \alpha_k^L | \hat{H}[k]_{\beta_k}^L |\alpha_k^{\prime L} \rangle \\
        =&\ \sum_{\alpha_{k-1} n_k;\alpha'_{k-1} n'_k}
         A[k]^{n_k}_{\alpha_{k-1}\alpha_k} A[k]^{n'_k}_{\alpha'_{k-1}\alpha'_k} \\
        &\ \quad \quad \quad \quad \quad \quad
        \times \langle \alpha_{k-1}^{L}\ n_k | \hat{H}[k]_{\beta_k}^{L} |\alpha_{k-1}^{\prime L}\ n'_k\rangle \\
    &\ \langle \alpha_k^R | \hat{H}[k]_{\beta_k}^R |\alpha_k^{\prime R} \rangle \\
        =&\ \sum_{n_{k+1}\alpha_{k+1};n'_{k+1} \alpha'_{k+1}}
         A[k+1]^{n_{k+1}}_{\alpha_k\alpha_{k+1}}
         A[k+1]^{n'_{k+1}}_{\alpha'_k\alpha'_{k+1}} \\
        &\ \quad \quad \quad \quad \quad \quad
        \times \langle n_{k+1}\ \alpha_{k+1}^R | \hat{H}[k]_{\beta_k}^{R}
        |n'_{k+1}\ \alpha_{k+1}^{\prime R} \rangle
    \label{eq:op-trans}
\end{aligned}
\end{equation}

\section{Notation for SU(2) spin-adapted \textit{ab initio} DMRG}
\label{app:qc-dmrg}

For the \textit{ab initio} DMRG implemented in this work, we associate each site \( k \) (\( k = 1,2,\cdots, K \)) with a spatial orbital. The \textit{ab initio} Hamiltonian is written as\cite{wouters2014density}
\begin{equation}
    \hat{H} = \sum_{ij,\sigma} t_{ij,\sigma} \ a_{i\sigma}^\dagger a_{j\sigma}
    + \frac{1}{2} \sum_{ijkl, \sigma\sigma'} v_{ijkl, \sigma\sigma'}\
    a_{i\sigma}^\dagger a_{k\sigma'}^\dagger a_{l\sigma'}a_{j\sigma}
    \label{eq:qc-hamil}
\end{equation}
where
\begin{equation}
\begin{aligned}
    t_{ij,\sigma} =&\ \int \mathrm{d}\mathbf{x} \
    \phi_{i\sigma}^*(\mathbf{x}) \left( -\frac{1}{2}\nabla^2 - \sum_a \frac{Z_a}{r_a} \right)
    \phi_{j\sigma}(\mathbf{x}) \\
    v_{ijkl,\sigma\sigma'} =&\
    \int \mathrm{d} \mathbf{x}_1 \mathrm{d} \mathbf{x}_2 \ \frac{\phi_{i\sigma}^*(\mathbf{x}_1)\phi_{k\sigma'}^*(\mathbf{x}_2)
    \phi_{l\sigma'}(\mathbf{x}_2)\phi_{j\sigma}(\mathbf{x}_1)}{r_{12}}
\end{aligned}
\end{equation}
with the following symmetry conditions
\begin{equation}
\begin{aligned}
    t_{ij,\sigma} =&\ t_{ji,\sigma} \\
    v_{ijkl, \sigma\sigma'} =&\ v_{jikl, \sigma\sigma'} = v_{ijlk, \sigma\sigma'}
        = v_{klij, \sigma'\sigma}
\end{aligned}
\end{equation}

With SU(2) spin symmetry we additionally have\cite{sharma2012spin}
\begin{equation}
\begin{aligned}
    t_{ij} =&\ t_{ij,\alpha} = t_{ij,\beta} \\
    v_{ijkl} =&\ v_{ijkl, \alpha\alpha} = v_{ijkl, \alpha\beta}
        = v_{ijkl, \beta\alpha} = v_{ijkl, \beta\beta}
\end{aligned}
\end{equation}

In conventional \textit{ab initio} DMRG, the left-right decomposition of the Hamiltonian (\autoref{eq:lr-decomp}) is written in terms of  \textit{normal} and \textit{complementary} operators.\cite{white1999ab} One can choose to use two-index complementary operators only with the right block (the Normal/Complementary (NC) partition) or only with the left block (the Complementary/Normal (CN) partition). The SU(2) spin-adapted left-right decomposition of the Hamiltonian using the NC and CN partition is respectively\cite{sharma2012spin}
\begin{multline}
    \hat{H}[k]^{\mathrm{NC}[0]} = \hat{H}^{L[0]} \otimes_{[0]} \hat{1}^{R[0]} + \hat{1}^{L[0]} \otimes_{[0]} \hat{H}^{R[0]}\\
        +2 \sum_{i\in L} \Big( a_i^{\dagger[\frac{1}{2}]} \otimes_{[0]} \hat{R}_i^{R[\frac{1}{2}]} + a_i^{[\frac{1}{2}]} \otimes_{[0]} \hat{R}_i^{R\dagger[\frac{1}{2}]} \Big) \\
        +2 \sum_{i\in R} \Big( \hat{R}_i^{L\dagger[\frac{1}{2}]} \otimes_{[0]} a_i^{[\frac{1}{2}]} + \hat{R}_i^{L[\frac{1}{2}]} \otimes_{[0]} a_i^{\dagger[\frac{1}{2}]} \Big) \\
        -\frac{1}{2} \sum_{ij\in L} \Big( \hat{A}_{ij}^{[0]} \otimes_{[0]} \hat{P}_{ij}^{R[0]} + \sqrt{3}\ \hat{A}_{ij}^{[1]} \otimes_{[0]} \hat{P}_{ij}^{R[1]} \\
        + \hat{A}_{ij}^{\dagger[0]} \otimes_{[0]} \hat{P}_{ij}^{R\dagger[0]}
        + \sqrt{3}\ \hat{A}_{ij}^{\dagger[1]} \otimes_{[0]} \hat{P}_{ij}^{R\dagger[1]}
        \Big) \\
        + \sum_{ij\in L} \Big( \hat{B}_{ij}^{[0]} \otimes_{[0]} \hat{Q}_{ij}^{R[0]}
        + \sqrt{3}\ \hat{B}_{ij}^{[1]} \otimes_{[0]} \hat{Q}_{ij}^{R[1]} \Big)
    \label{eq:h-nc}
\end{multline}
and
\begin{multline}
    \hat{H}[k]^{\mathrm{CN}[0]} = \hat{H}^{L[0]} \otimes_{[0]} \hat{1}^{R[0]} + \hat{1}^{L[0]} \otimes_{[0]} \hat{H}^{R[0]} \\
        +2 \sum_{i\in L} \Big( a_i^{\dagger[\frac{1}{2}]} \otimes_{[0]} \hat{R}_i^{R[\frac{1}{2}]} + a_i^{[\frac{1}{2}]} \otimes_{[0]} \hat{R}_i^{R\dagger[\frac{1}{2}]} \Big) \\
        +2 \sum_{i\in R} \Big( \hat{R}_i^{L\dagger[\frac{1}{2}]} \otimes_{[0]} a_i^{[\frac{1}{2}]} + \hat{R}_i^{L[\frac{1}{2}]} \otimes_{[0]} a_i^{\dagger[\frac{1}{2}]} \Big) \\
        -\frac{1}{2}  \sum_{ij\in R} \Big( \hat{P}_{ij}^{L[0]} \otimes_{[0]} \hat{A}_{ij}^{[0]} + \sqrt{3}\ \hat{P}_{ij}^{L[1]} \otimes_{[0]} \hat{A}_{ij}^{[1]} \\
        + \hat{P}_{ij}^{L\dagger[0]} \otimes_{[0]} \hat{A}_{ij}^{\dagger[0]}
        + \sqrt{3} \ \hat{P}_{ij}^{L\dagger[1]} \otimes_{[0]} \hat{A}_{ij}^{\dagger[1]}
        \Big) \\
        + \sum_{ij\in R} \Big( \hat{Q}_{ij}^{L[0]} \otimes_{[0]}
        \hat{B}_{ij}^{[0]} + \sqrt{3}\ \hat{Q}_{ij}^{L[1]} \otimes_{[0]}
        \hat{B}_{ij}^{[1]} \Big)
    \label{eq:h-cn}
\end{multline}
where the superscript and subscript  \( [S] \) are used to indicate the total spin quantum number for the spin tensor operator and the resulting spin tensor operator obtained from the tensor product, respectively, and the block Hamiltonian \( \hat{H}^{L/R[0]} \), normal operators \( \hat{A}_{ij}^{[S]} \), \( \hat{B}_{ij}^{[S]} \), and complementary operators \( \hat{R}_i^{L/R[\frac{1}{2}]}\), \( \hat{P}_{ij}^{L/R[S]}\), \( \hat{Q}_{ij}^{L/R[S]} \) are defined by
\begin{equation}
\begin{aligned}
    \hat{R}_i^{L/R[\frac{1}{2}]} =&\ \frac{\sqrt{2}}{4} \sum_{j\in L/R} t_{ij}\  a_j^{[\frac{1}{2}]} \\
        &\ \quad + \sum_{jkl\in L/R} v_{ijkl} \Big( a_{k}^{\dagger[\frac{1}{2}]} \otimes_{[0]} a_l^{[\frac{1}{2}]} \Big) \otimes_{[\frac{1}{2}]} a_j^{[\frac{1}{2}]}, \\
    \hat{A}_{ij}^{[0/1]} =&\ a_i^{\dagger[\frac{1}{2}]} \otimes_{[0/1]} a_j^{\dagger[\frac{1}{2}]}, \\
    \hat{B}_{ij}^{[0/1]} =&\ a_i^{\dagger[\frac{1}{2}]} \otimes_{[0/1]} a_j^{[\frac{1}{2}]}, \\
    \hat{P}_{ik}^{L/R[0/1]} =&\ \sum_{jl\in L/R} v_{ijkl}\ a_j^{[\frac{1}{2}]} \otimes_{[0/1]} a_l^{[\frac{1}{2}]}, \\
    \hat{Q}_{ij}^{L/R[0]} =&\ \sum_{kl\in L/R} \big( 2v_{ijkl} - v_{ilkj} \big)
        a_k^{\dagger[\frac{1}{2}]} \otimes_{[0]} a_l^{[\frac{1}{2}]}, \\
    \hat{Q}_{ij}^{L/R[1]} =&\ \sum_{kl\in L/R} v_{ilkj}\ 
        a_k^{\dagger[\frac{1}{2}]} \otimes_{[1]} a_l^{[\frac{1}{2}]}
    \label{eq:op-def}
\end{aligned}
\end{equation}
with the following symmetry conditions (when \( i \neq j \))\cite{kurashige2009high}
\begin{equation}
\begin{aligned}
    \hat{A}_{ij}^{[S]} =&\ (-1)^S \hat{A}_{ji}^{[S]} \\
    \hat{B}_{ij}^{[S]} =&\ (-1)^S \big( \hat{B}_{ji}^{[S]} \big)^\dagger \\
    \hat{P}_{ij}^{[S]} =&\ (-1)^S \hat{P}_{ji}^{[S]} \\
    \hat{Q}_{ij}^{[S]} =&\ (-1)^S \big( \hat{Q}_{ji}^{[S]} \big)^\dagger
    \label{eq:index-sym}
\end{aligned}
\end{equation}

The corresponding MPO for the NC and CN partitions can be constructed based on the blocking formulae for the spin tensor operators, and \autoref{eq:h-nc} and \autoref{eq:h-cn}, respectively. Since these operators have at most two spatial orbital indices, the MPO bond dimension \( D \sim K^2 \). The blocking formulae explicitly yield only the non-zero elements of the MPO, and thus using the blocking formulae in the DMRG algorithm can be viewed as implementing sparse tensor contraction with the MPO.

Alternatively, there are procedures to automatically construct the elements of the MPO tensors by matrix decomposition (and other algorithms) simply given the list of two-electron integrals. 
Examples of these automated MPO construction approaches are the fork-merge approach,\cite{keller2015efficient} the SVD approach,\cite{chan2016matrix} the delinearization approach,\cite{hubig2017generic} and the bipartite approach.\cite{ren2020general} Note that some of these procedures  work with a dense matrix representation of the MPO tensors (even if the matrices have exact zeros). As discussed in the main text, the sum of sub-Hamiltonians allows for the correct scaling of implementations which use such MPO construction techniques without explicit implementation of sparse tensor algebra. Thus the strategies in this work,
when used with 
automated MPO construction techniques, achieve both the correct serial cost as well as have a low communication overhead for parallel scaling.

\bibliography{para-dmrg}

\end{document}